\documentstyle[amssymb,aps,multicol,epsfig]{revtex}

\begin{document}
\title{Colossal Magnetoresistance is a Griffiths Singularity}
\author{M. B. Salamon, P. Lin}
\address{Department of Physics\\
University of Illinois\\
1110 West Green Street, Urbana, IL 61801-3080}
\author{S. H. Chun}
\address{Department of Physics\\
Pennylvania State University\\
University Park, PA 16802-6300}
\date{\today }
\maketitle

\begin{abstract}
It is now widely accepted that the magnetic transition in doped manganites
that show large magnetoresistance is a type of percolation effect. This
paper demonstrates that the transition should be viewed in the context of
the Griffiths phase that arises when disorder suppresses a magnetic
transition. \ This approach explains unusual aspects of susceptibility and
heat capacity data from a single crystal of La$_{0.7}$Ca$_{0.3}$MnO$_{3}.$
\end{abstract}


\-%
\begin{multicols}{2}%
%
The term ``colossal magnetoresistance'' (CMR) has commonly been used to
describe the very large, magnetic-field driven changes in electrical
resistivity in oxides based on LaMnO$_{3}$ near their second-order,
ferromagnetic transitions. The largest CMR effects are accompanied by other
anomalies in magnetic and thermodynamic properties. \ Among these are the
failure of the magnetic correlation length to increase strongly as the
transition temperature $T_{C}$ is approached from above, the persistence of
a well defined spin-wave dispersion close to the transition \cite{lynn96},
and an unusual shift in the heat capacity peak to higher temperatures in
applied magnetic fields.\cite{lin00} An explanation of this unusual behavior
of the heat capacity in the context of a Griffiths singularity \cite
{griffiths69} is the focus of the present paper.

There is now general consensus that the CMR transition is a type of
percolation in which, due to the double-exchange process \cite{zener51},
bonds become metallic as neighboring spins tend to align. The strength of
the CMR\ effect (along with the transition temperature) depends strongly on
the ionic size and concentration of the divalent atom that substitutes for
La in LaMnO$_{3}$.\ The effect is nearly absent for La$_{2/3}$Sr$_{1/3}$MnO$%
_{3}$ ($T_{C}$ = 360\ K) but quite strong in the present sample La$_{0.7}$Ca$%
_{0.3}$MnO$_{3}$ ($T_{C}=218$ K) \cite{okuda00}.\ In the low temperature
metallic phase, the exchange interactions, as measured by the spin-wave
stiffness, are the same in Sr- snd Ca-doped crystals \cite{rhyne00},
demonstrating that the key to the CMR effect is to be found in the
non-metallic regime. The ionic size of the divalent substituent exerts its
effect on magnetic and electronic properties through local tilting of the
oxygen octahedra \cite{garciamunoz96} and the concurrent bending of the
active Mn-O-Mn bonds. This inhibits the formation of metallic bonds and
leads to charge localization, polaron formation, and possible charge
segregation \cite{coey99,dagotto01}. Evidence in favor of a percolation
picture comes from a Monte-Carlo simulation of a random field Ising model
that assigns conductivity zero and unity to bonds between neighboring
antiparallel and parallel spins respectively \cite{mayr01}. It both produces
CMR\ and emphasizes the importance of randomness. Experimental evidence was
provided by Jaime et al. \cite{jaime99a} who extracted the field and
temperature dependent metallic-bond concentration $c(H,T)$ from the
resistivity via the effective medium approximation, and showed that it also
describes the thermoelectric power data. Direct evidence for coexisting
polaronic/insulating and metallic components have been reported from neutron 
\cite{dai00a}, \cite{adams00}, electron \cite{zuo01} and Raman scattering 
\cite{yoon98} studies.

Discussions of the CMR\ effect in percolation terms \cite
{gorkov98a,podzorov00,raquet00} have generally neglected a central point:
that the percolating entities are formed {\em thermodynamically} as the
temperature is lowered, nucleated by the intrinsic randomness of a doped
material and amplified by the tendency for polaron formation and charge
segregation \cite{arovas99}. Griffiths \cite{griffiths69} first pointed out,
in the context of dilution (bonds randomly assigned values of 0 or $J$),
that singularities arise in thermodynamic properties in the temperature
range $T_{C}^{%
\mathop{\rm rand}%
}\leq T\leq T_{G}$ between the random transition $T_{C}^{%
\mathop{\rm rand}%
}$ and the ``pure'' transition temperature $T_{G}.$ Bray \cite{bray82,bray87}
extended the argument to any bond distribution that reduces the transition
temperature, terming the regime between $T_{C}^{%
\mathop{\rm rand}%
}$ and $T_{G}$ the ``Griffiths phase.''\ In this paper, we use the Bray
model to demonstrate that the zero-field transition is an example of a
Griffiths singularity. To treat the field dependence, we consider two
possible exchange energies: $J_{met},$ associated with the double-exchange
process on metallic Mn-O-Mn bonds, and $J_{ins},$ the residual ferromagnetic
interaction that exists on insulating bonds. The proportion of metallic
bonds changes with field and temperature, monitored through the
resistivity-based metallic fraction $c(H,T).$\ \ 

Among the effects that characterize the Griffiths phase are a distribution
of susceptibilities and slow spin dynamics. \ The latter have been observed
by muon relaxation \cite{heffner97b,heffner99} and show up strongly in noise
measurements \cite{merithew00}. \ We focus in this work on the nature of the
susceptibility and magnetization, and on the behavior of the heat capacity
in applied magnetic fields, tying the last to the CMR effect itself. \
Figure 1 shows the inverse susceptibility $\chi ^{-1}(T)=H/M(T)$ of a single
crystal sample of La$_{0.7}$Ca$_{0.3}$MnO$_{3},$ where $M(T)$\ is the
magnetization in a field of 1 kOe. \ The high-temperature behavior is of
Curie-Weiss form, with a paramagnetic Curie temperature $\Theta =202$ K.
Well above $T_{C}=218$ K the slope of the experimental curve corresponds to $%
S\approx 8$, much shallower than the Curie-Weiss law for spin $S_{eff}=1.85$
(the weighted average of $S=3/2$ \ and $S=2$ appropriate for this sample),
shown as a dashed line. Further, there is a sharp downturn in $\chi ^{-1}(T)$
before the Curie temperature is reached. This alone identifies the
transition as a Griffiths singularity, characterized by a susceptibility
exponent less than unity \cite{castroneto98}; that is, $\chi ^{-1}(T)\propto
(T-T_{C}^{%
\mathop{\rm rand}%
})^{1-\lambda },$ with a non-universal \ positive exponent $\lambda \leq 1.$

In his treatment of the Griffiths phase, Bray \ \cite{bray82,bray87} argues
that the distribution of cluster sizes can be characterized by the
longitudinal susceptibility matrix $\chi _{ij}=g^{2}\mu _{B}^{2}\left\langle
S_{i}^{z}S_{j}^{z}\right\rangle /k_{B}T$ or rather by its inverse, whose
eigenvalues $\mu /C$ are distributed according to $p(\mu ),$ where

\begin{equation}
p(\mu )\varpropto \mu ^{-\gamma }e^{-A(T)/\mu }.  \label{eq1}
\end{equation}
Here $C$ is the Curie constant of the material and each $\mu $ represents an
effective Curie-Weiss temperature difference \ $T-\Theta _{eff}(T)$. The $%
\Theta _{eff}(T)$ may be equally well related \cite{bray82} to the
eigenvalue distribution of the exchange matrix which the double-exchange
mechanism makes temperature and field dependent. The close connection
between the effective exchange energy and the metallicity of the
corresponding Mn-O-Mn bond ties the Griffiths behavior to the CMR\ effect.
Bray ignored the negative-power prefactor, but it is required if $p(\mu )$
is to decrease for large $\mu $ and to rise sharply near $\mu =0$ as $T_{C}^{%
\mathop{\rm rand}%
}$ is approached \cite{bray82}. \ As all clusters are finite, none is fully
ferromagnetic so we must have $p(\mu \rightarrow 0)=0$. However, there is a
pile-up of large clusters as the transition is approached, leading Bray to
deduce that $A(T)$ vanishes as, 
\begin{equation}
A(T)=T_{0}\left( \frac{T-T_{C}^{%
\mathop{\rm rand}%
}}{T_{C}^{%
\mathop{\rm rand}%
}}\right) ^{2(1-\beta )},  \label{eq2}
\end{equation}
where $T_{0}$ is a parameter and $\beta =0.38$ is the order-parameter
exponent for the pure system, assumed to be $3D$ Heisenberg-like. Because $%
\mu =T$ \ for free spins, we cut off the distribution $p(\mu )$ and
calculate the average inverse susceptibility $\left\langle \mu \right\rangle
/C$, using 
\begin{equation}
\left\langle \mu \right\rangle =\frac{\int_{0}^{T}\mu p(\mu )d\mu }{%
\int_{0}^{T}p(\mu )d\mu }  \label{eq3}
\end{equation}
\newline
The resul is shown as a solid line in Fig. 1, where the Curie constant is 
\begin{equation}
C=\frac{Ng^{2}\mu _{B}^{2}S_{eff}(S_{eff}+1)}{3k_{B}}=0.074\text{ K}
\label{eq4}
\end{equation}
and $T_{C}^{%
\mathop{\rm rand}%
}=$ 223 K. The free parameters are the characteristic temperature $T_{0}=9.5$
K \ and the non-universal prefactor exponent, $\gamma =1.59\pm 0.5.$ \ The
calculated curve is very close to the power law $\chi ^{-1}(T)\propto
(T-T_{C}^{%
\mathop{\rm rand}%
})^{0.72},$ so that $\lambda =0.28,$ as seen in the inset to Fig. 1. The
values of $T_{0}$ and $\gamma $ are strongly correlated; the data can be fit
with values of $\gamma $ in the range cited, resulting in values of $\lambda 
$ in the range $0.2\leq \lambda \leq 0.3.$ \ Analysis of the quantum
critical point of UCu$_{4}$Pd in terms of a Griffiths singularity yields the
much larger value $\lambda =2/3$ \cite{castroneto98}.

The downturn in the $\chi ^{-1}(T)$ data close to the transition temperature
is more abrupt than Bray's approach predicts and $T_{C}^{%
\mathop{\rm rand}%
}$\ is somewhat higher than $T_{C}=$218.2 K, the location of the zero-field
heat capacity peak shown in Fig. 2 \cite{lin00} that we take to mark the
true transition temperature. Griffiths, in his original paper \cite
{griffiths69}, suggested that the susceptibility would tend to diverge in
advance of the onset of true long range order, and that may explain the
present result. \ The nonanalytic behavior of the magnetization at small
magnetic fields is readily observable in the vicinity of $T_{C}$ where the
magnetization rises so abruptly that, considered in terms of conventional
critical behavior $M\propto H^{1/\delta },$ it requires $\delta \approx 13,$
far from any standard value$.$\ Magnetization data taken at 218 K, showing
this rapid rise, are shown in Fig. 3.\ The upward curvature in the $\chi
^{-1}(T)$ data at high temperatures has also been reported by, for example,
de Teresa et al.\cite{deteresa97a}, who find a linear regime at high
temperatures with a slope consistent with the Curie constant in Eq.(\ref{eq4}%
) and an effective Curie-Weiss temperature $\approx 1.25T_{C}$; it is
tempting to assign this to $T_{G}$.\ 

Figure 2 shows the heat capacity of the sample in fields of 0 T, 3 T and 7 T
after subtraction of a smooth background that fits the data at temperatures
well away from the peak. The sharpness of the zero field data might signal a
first-order transition, but there is no evidence of hysteresis. The shift in
the heat capacity peak is uncharacteristic of ferromagnetic transitions, as
discussed previously \cite{lin00}. To treat this behavior, we employ a
modified version of the \ Oguchi cluster method \cite{smart}, a simple
improvement to mean-field theory. In this method, the energy of a pair is
treated exactly, while coupling to its $z-1$ neighbors is handled in
mean-field theory. \ The magnetic part of the double-exchange energy is $%
xt[\cos (\theta /2)-\overline{\cos (\theta /2)}],$ where $\theta $ is the
classical angle between core spins,\ $t$ is the transfer energy, $x$ is the
doping level and the bar denotes the angular average. For the Oguchi
calculation, we use the quantum analog \cite{anderson55} 
\begin{equation}
E_{de}(S_{0})=xt\left[ \frac{S_{0}+1/2}{2S+1}-\overline{\frac{S_{0}+1/2}{2S+1%
}}\right] ,  \label{eq0}
\end{equation}
where $S_{0},$ the total spin of the two manganese $S_{c}=3/2$ cores and the
shared $e_{g}$ electron, is in the range $1/2\leq S_{0}\leq 7/2$. The pair
is coupled to the magnetization of its $z-1$ neighbors via either metallic $%
J_{met}$ or insulating $J_{ins}$ exchange bonds. Assuming spherical clusters
(of undetermined size), we follow the effective-medium approach of Jaime, et
al. \cite{jaime99a} and extract the metallic bond concentration $c(H,T)$
from the measured resistivity $\rho (H,T)$ (Fig. 2a) by extrapolating the
fitted the high-temperature (activated) and low temperature (power-law)
resistivities. The resulting $c(H,T)$ is shown for several applied fields in
Fig. 2b.\ The effective field acting on the pair is, then, 
\begin{equation}
H_{eff}=H+2(z-1)S_{eff}J_{eff}m(H,T)/g\mu _{B},
\end{equation}
where $m(H,T)$ is the reduced magnetization, to be calculated
self-consistently, and $J_{eff}=c(H,T)J_{met}+[1-c(H,T)]J_{ins}$. The
resultant $m(H,T)$ values are then used to determine the pair energy $%
\left\langle E_{de}(S_{0})\right\rangle $ within the Oguchi approach.

The field and temperature dependent heat capacity calculated from $%
\left\langle E_{de}(S_{0})\right\rangle $ with a single amplitude parameter,
shown as solid lines in Fig. 2c, is in good agreement with the data. The
insulating exchange energy is determined from the observed Curie constant to
be $J_{ins}=3k_{B}\Theta /2zS_{eff}(S_{eff}+1)=0.82$ meV. We obtain the
transfer energy $t$ from the spin-wave stiffness $D$ according to $t\approx
(2S_{c}+1)D/xa^{2}$ \cite{ohata73}. The energy difference between parallel
and antiparallel configurations is $E_{de}(7/2)-E_{de}(1/2)=3xt/5$; for a
Heisenberg interaction $-2J_{met}\overrightarrow{S}_{1}\cdot \overrightarrow{%
S}_{2}$ the difference would be $4S^{2}J_{met}$ giving $J_{met}=3xt/20S^{2}.$
For $D=150$ mev \AA $^{2},$ which is common to most manganites at this
doping level \cite{rhyne00}, we find $t=130$ meV and $J_{met}=1.46$ meV.
Clearly, the background subtracted from the data has ignored magnetic
contributions that remain well below $T_{C}$. The shoulder in the zero-field
curve arises from the tail in $c(0,T)$ in the vicinity of $T_{C}.$ This may
signal a break-down in the effective medium approximation, as very large
clusters are known from noise measurements \cite{merithew00,hess01} to have
a dramatic effect on the zero-field resisitivity near the transition. \ The
self-consistent values $m(H,18$ K), shown as open circles in the inset of
Fig. 3, match the measured critical isotherm. Neither $J_{met}$ nor $J_{ins}$
determines the location of the zero-field heat capacity peak at 218.2 K. \
Rather, the rapid buildup in both the size and number of magnetic clusters
that give rise to the Griffiths behavior of the susceptibility causes a
concomitant increase in the conductivity of the material. As in Griffiths'
dilution picture, the existence of $J_{met}$ serves to pull the transition
temperature upward from the values expected from $J_{ins}.$ Insofar as it is
characterized by the spin-wave dispersion at low temperatures, $J_{met}$ is
essentially the same for all CMR materials. \ Therefore, the stronger
tendency toward self-trapping that occurs as the Mn-O-Mn bond angle is
reduced, lowers $J_{ins}$ relative to $J_{met}$ and increases the dominance
of the Griffiths phase.

We have argued that the CMR effect should be treated in the context of a
Griffiths singularity driven by intrinsic randomness, the combined effect of
doping, the tendency for charge segregation, and the self-trapping effect
associated with polaron formation. \ Bray's treatment of the distribution of
susceptibilities explains the unusual downturn in the measured
susceptibility and the magnitude of the effective paramagnetic moment. \
More importantly, we have used the Griffiths idea of a distribution of
exchange energies to demonstrate that the magnetic transition is driven by
the accumulation of larger exchange energies on metallic bonds as the system
begins to order. Magnetic fields align large clusters, turning bonds along
their shared boundaries from insulating to metallic. This has the effect of
shifting the effective percolation point upward in temperature, and
connecting the conductive network at temperatures well above the zero-field
percolation point. This is the essence of the CMR effect. A modified Oguchi
model that treats pairs in the context of the double-exchange mechanism, and
that uses the resistivity-derived, metallic-bond concentration, accurately
captures the nature of short range correlations, although not the effect of
larger clusters. Future work will address the self-consistent calculation of
the concentration and magnetization without the need for empirical input
from the CMR data, and will apply this analysis to other CMR transitions.

We gratefully acknowledge the contribution of Y. Tomioka and Y. Tokura in
providing the sample used for this study and have benefitted from
discussions with M. B. Weissman. This work was supported in part by the U.S.
Department of Energy through contract DEFG02-91ER45439 in the F. Seitz
Materials Research Laboratory.

\begin{figure}
\psfig{figure=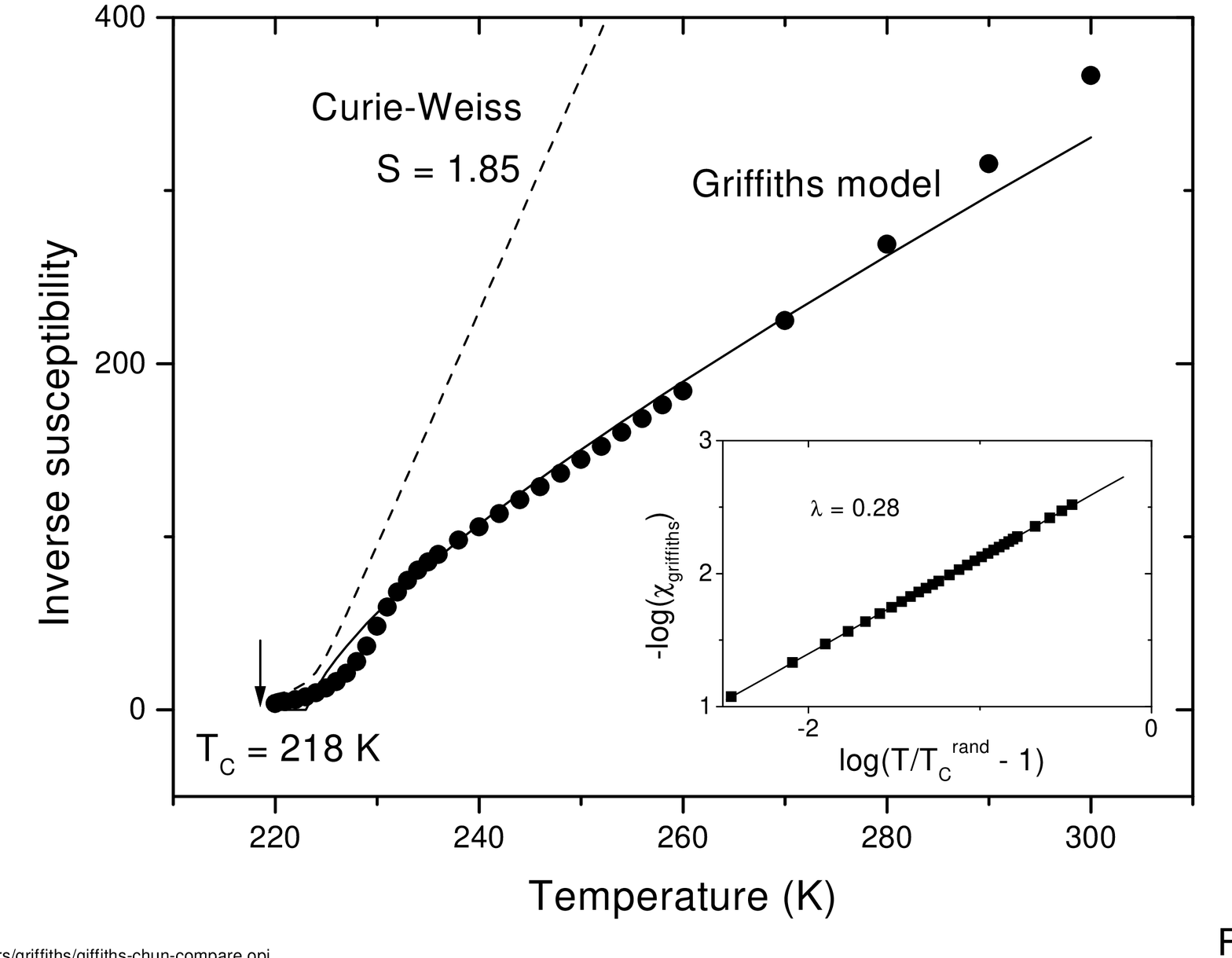,width=8cm,clip=}
\noindent
\caption{Ratio of the magnetic field $H$ to the magnetization $M$ at 1 kOe. The dashed 
line is the expected Curie-Weiss behavior in a 1 kOe field for spin $S = 1.85$ and 
$T_C^{rand}=223$ K while the solid line is the result of the Griffiths-phase, $H=0$ 
calculation. The inset shows the Griffiths phase 
result vs reduced temperature on a log-log plot, demonstrating its power-law behavior.}
\label{fi1}
\end{figure}%
%

\begin{figure}
\psfig{figure=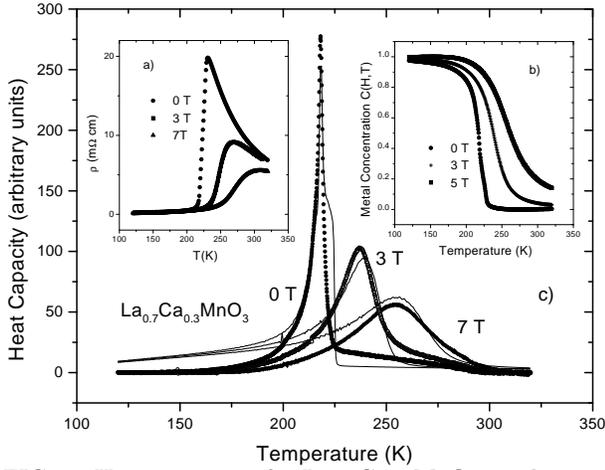,width=8cm,clip=}
\noindent
\caption{Heat capacity of a La$_{0.7}$Ca$_{0.3}$MnO$_3$ single crystal at several 
fields. The solid lines are computed from the measured metallic-bond fraction using the
Oguchi model.  The  resistivity and metallic fractions at the same fields are shown in
insets a) and b) respectively.}
\label{fig2}
\end{figure}%
%

\begin{figure}
\psfig{figure=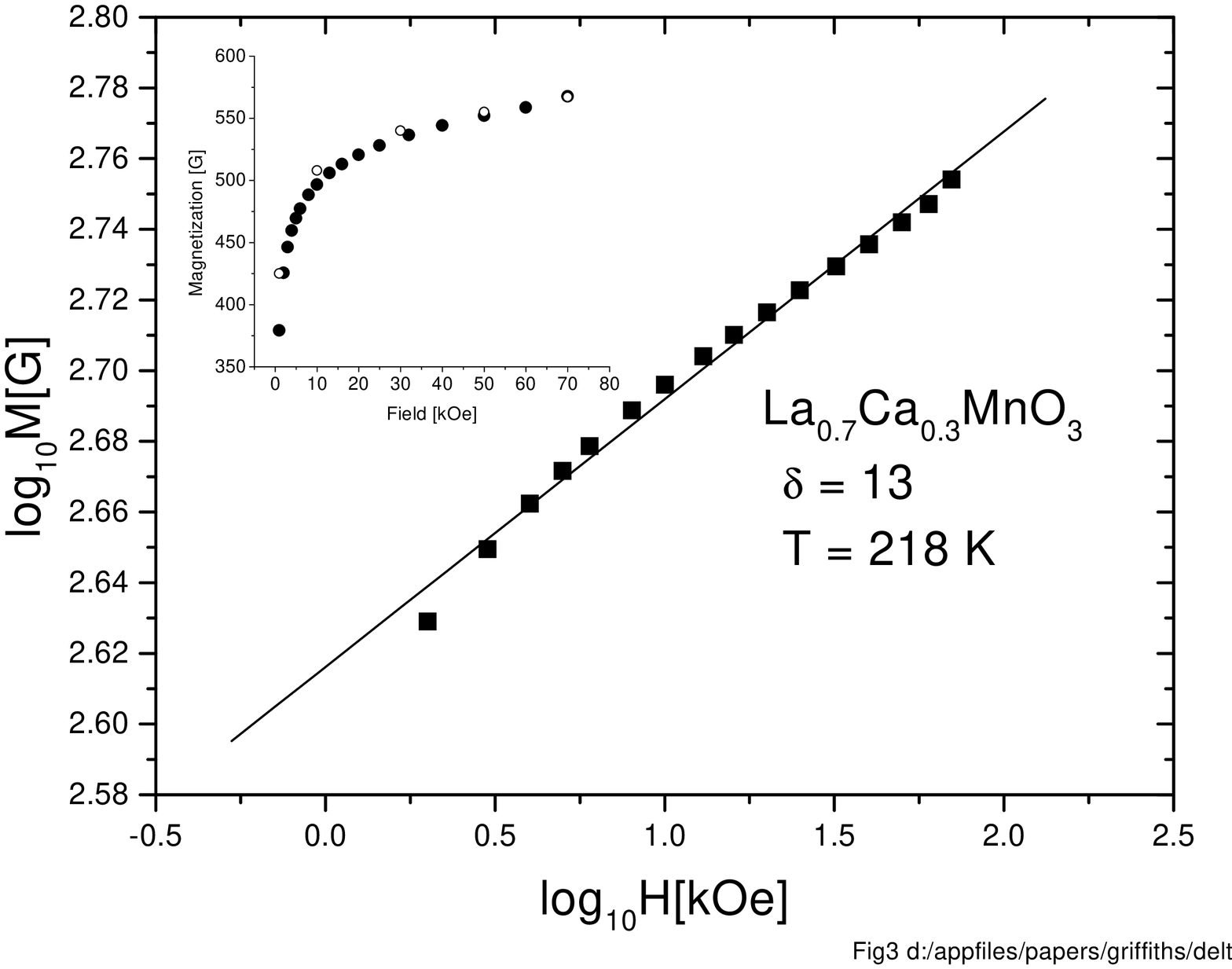,width=8cm,clip=}
\noindent
\caption{Magnetization vs applied field at 218 K. The inset is a linear plot; the main panel
is logarithmic, showing the small slope given by $1/\delta$. The open circles in the inset
are calculated at 218 K within the Oguchi model.}
\label{fig3}
\end{figure}%
%

\bibliographystyle{prsty}
\bibliography{bibcmr}

\end{multicols}%
%

\end{document}